# Fog Networking: An Overview on Research Opportunities

Mung Chiang
Arthur LeGrand Doty Professor of Electrical Engineering
Princeton University

December 2015

## A. Introduction

The past 15 years have seen the rise of the Cloud, along with rapid increase in Internet backbone traffic and more sophisticated cellular core networks. There are three different types of "Clouds:" (1) data center, (2) backbone IP network and (3) cellular core network, responsible for computation, storage, communication and network management. Now the functions of these three types of Clouds are "descending" to be among or near the end users, i.e., to the edge of networks, as "Fog."

We take the following as a working definition of Fog Networks: "*It is an architecture that users one or a collaborative multitude of end-user clients or near-user edge devices to carry out a substantial amount of storage, communication and management.*" Architecture allocates functionalities. Engineering artifacts that may use a Fog architecture include 5G, home/personal networking, and the Internet of Things.

To highlight the contrast between Cloud and Fog, we can compare Fog architecture with the current standard practice along the following three dimensions:

- Carry out a substantial amount of *storage* at or near the end-user (rather than stored primarily in large-scale data centers).
- Carry out a substantial amount of *communication* at or near the end-user (rather than all routed through the backbone network).
- Carry out a substantial amount of *management*, including network measurement, control and configuration, at or near the end-user (rather than controlled primarily be gateways such as those in the LTE Core).

It is not a binary choice between Cloud and Fog either: they form a mutually beneficial, inter-dependent continuum. It is a continuum: to the wearable devices, a mobile phone may be viewed as the Cloud. They are inter-dependent, e.g., coordination among devices in a Fog may rely on the Cloud. They are also mutually beneficial: certain functions are naturally more advantageous to carry out in Fog while others in Cloud. The interface between Cloud and Fog is indeed a key aspect of Fog R&D.

Fog architectures may be fully distributed, mostly centralized or somewhere in-between. They may rely on hardware, software, or combination of both. The common denominator is that they distribute the resources and services of computation, communication, control, and storage closer to devices and systems at or near the users.

There is already a large and increasing range of such client and edge devices today: from smart phones to tablets and from set-top boxes to small cell base stations. Some of them have become dramatically more powerful in computation, communication, storage and sensing capabilities within the past several years, while still limited in other ways such as energy supply. As different segments of Internet of Things (IoT), Internet of Everything (IoE) or Internet of Me (IoM) start to take off, e.g., consumer, wearable, industrial, enterprise, automobile, healthcare, building, energy, etc., there will be an even more impressive surge in the diversity, volume, and capabilities of such "Fog nodes." Indeed, Fog's first application contexts was connected cars as proposed several years ago [1]. Information-transmitting light-bulbs, computers on a stick, and button-sized RF tuners further underscore the opportunities of Fog nodes.

Concurrently, wireless networks are increasingly used locally, e.g., within connected cars, smart-buildings, and personal body-area networks; and data generated locally is increasingly consumed locally. What can a crowd of such devices collectively accomplish, through a dense, distributed and under-organized network on the edge? What can they accomplish to enable ultra-low and deterministic latency, data mining in real time with streaming data, and cyber physical network's actuation and control functions within stringent temporal constraints?

It has become both feasible and interesting to ask the question: "What can be done on the network edge?" For example, what is the set-top box in your living room replaces the deep inspection boxes in operator network? And the dashboard in your car is your primary caching device? What if your phone and phones of others collectively act as controller similar to an S-GW or PDN-GW? While the answers to questions such as these may not be positive in all cases, it has finally become worthwhile asking the questions.

We may contrast these clients, edge devices, and "things" with the large, expensive, hard-to-innovate "boxes" in the Cloud: S-GW and PDN-GW in LTE core, large servers and switches inside a data center, and metro and core routers in wide-area-network backbone. The traditional view is that edge uses the core network and data centers. The Fog view is that edge *is* the core network and a data center. In the tension between the "brick" versus the "click," the pendulum is starting to swing back toward the "brick," where physical interactions with the cyber-system is becoming once again important.

**B. Why Fog and Why Now?**

Why would we be interested in the Fog view now? There are four main reasons:

1. ***Time:*** *Real time processing and cyber-physical system control.* Edge data analytics, as well as the actions it enables through control loops, often have stringent time requirement and can only be carried out on the edge, "here and now." This is particularly essential for Tactile Internet: the vision of millisecond reaction time on networks that enable virtual-reality-type interfaces between humans and devices.

2. **Cognition:** *Awareness of Client-centric objectives*. Following the end-to-end principle, some of the applications can be best enabled by knowing the requirements on the clients. This is especially true when privacy and reliability cannot be trusted in the Cloud, or when security is enhanced by shortening the extent over which communication is carried out.

3. **Efficiency:** *Pooling of local resources*. There are typically hundreds of gigabytes sitting idle on tablets, laptops and set-top boxes in a household every evening, or across a table in a conference room, or among the passengers of a public transit system. Similarly, idle processing power, sensing ability and wireless connectivity within the edge may be pooled within a Fog network.

4. **Agility:** *Rapid innovation and affordable scaling*. It is usually much faster and cheaper to experiment with client and edge devices. Rather than waiting for vendors of large boxes inside the network to adopt an innovation, in the Fog world a small team may take advantages of smart phone API and SDK, proliferation of mobile apps, and offer a networking service through its own API.

There are also two more "defensive" reasons for the rise of the Fog:

*Feasibility to operate on encrypted and multipath traffic*. A major trend these days is that by the time traffic leaves the edge and enters the backbone network, it is already encrypted and possibly traversing multiple paths, making it expensive if not impossible to operate on such data.

*In United States, implications of FCC Title II Ruling.* The FCC vote in February 2015 to classify Internet services, including mobile services, as a "utility" under Title II regulatory mandate, may further push network innovation to the edge in the US. A new regulatory environment does not mean networks cannot be engineered and managed anymore, but we may need different vantage points of control: not from inside the network but from around the end users. For example, today network operators can pick which lane (WiFi, Macro-cellular, Femtocell) a user device should be in. Since different lanes have different speeds and different payment system/amount, such practice may not be allowed any more in the US. Instead, we need to better design system where each user device must choose which lane to be in for itself. The challenge resulting from Title II regulation is a "hanging sword" that chills the deployment of network infrastructure innovations, as risk-Return balance now tips towards "keep the network as is." However, as long as the government does not prohibit end-user choices, then we can run networking from the edge, through client/home-driven control/configuration.

**C. Case Studies**

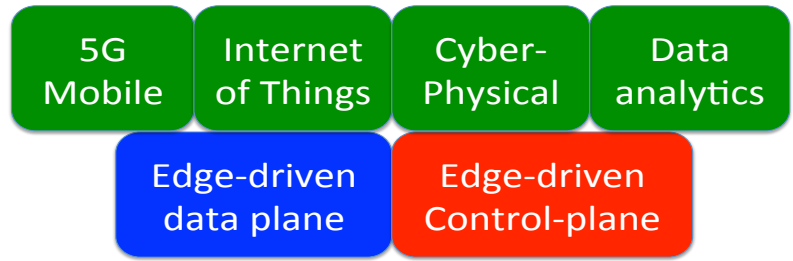

Figure 1: Data plane and control plane of Fog networks enable different applications

Architectural R&D asks the question of "who does what, at what timescale, and how to put the modules back together?" As an architecture, Fog networking consists of both data plane and control plane, each with a rapidly growing number of examples across protocol layers from the physical layer to the application layer:

- *Examples of Data plane of Fog:*
    o Pooling of clients idle computing/storage/bandwidth resources and local content
    o Content caching at the edge and bandwidth management at home
    o Client-driven distributed beam-forming
    o Client-to-client direct communications (e.g., FlashLinQ, LTE Direct, WiFi Direct, Air Drop)
    o Cloudlets and micro data-centers

- *Examples of Control plane of Fog:*
    o Over the Top (OTT) content management
    o Fog-RAN: Fog driven radio access network
    o Client-based HetNets control
    o Client-controlled Cloud storage
    o Session management and signaling load at the edge
    o Crowd-sensing inference of network states
    o Edge analytics and real-time stream-mining

Data-plane of Fog has been more extensively studied, e.g., [2]. In the following, we highlight a few particular cases that illustrate the potential and challenges of Fog control plane, such as the inference, control, configuration and management of networks:

*Case 1: Crowd-sensing LTE states (in commercial deployment).* Through a combination of passive measurement (e.g., RSRQ), active probing (e.g., packet train), application throughput correlation and historical data mining, a collection of client devices may be able to, in real-time and useful accuracy, infer the states of an eNB such as the number of Resource Blocks used [3].

*Case 2: OTT network provisioning and content management (in commercial deployment).* The traditional approach to innovating networks is to introduce another box inside the network, possibly a virtualized box but a box nonetheless. Fog networking directly leverages the "things" and phones instead, and removes the dependence on boxes-in-the-network altogether. With SDKs sitting behind apps on client devices, through tasks such as URL wrapping, content tagging, location tracking, behavior monitoring, network services can be innovated much faster.

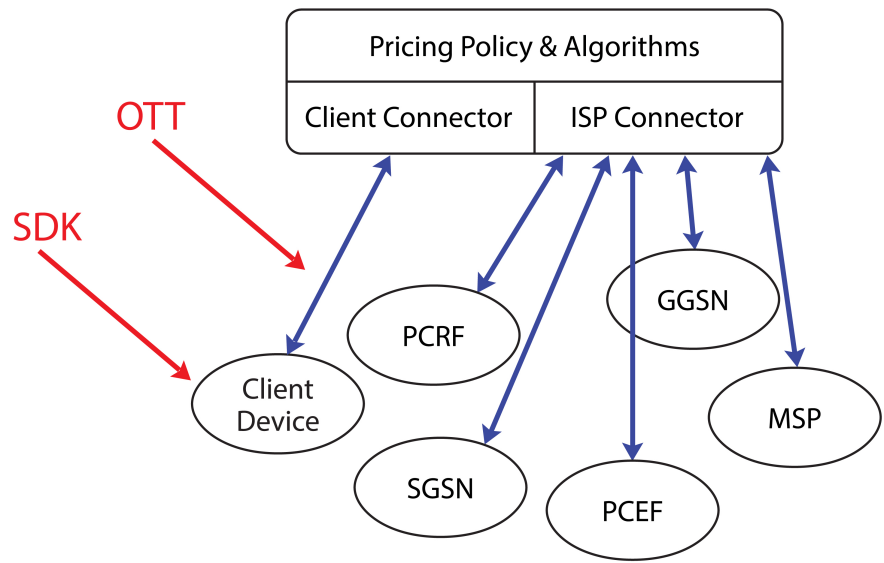

Figure 2: SDK sitting inside clients can enable network inference and configuration

*Case 3: Client-based HetNets control (in 3GPP standards).* Coexistence of heterogeneous networks (e.g., LTE, femto, WiFi) coexistence is a key feature in cellular networks today. Rather than through network operator control, each client can observe its local conditions and make decision on which network to join. Through randomization and hysteresis, such local actions may emerge globally to converge to a desirable configuration [4].

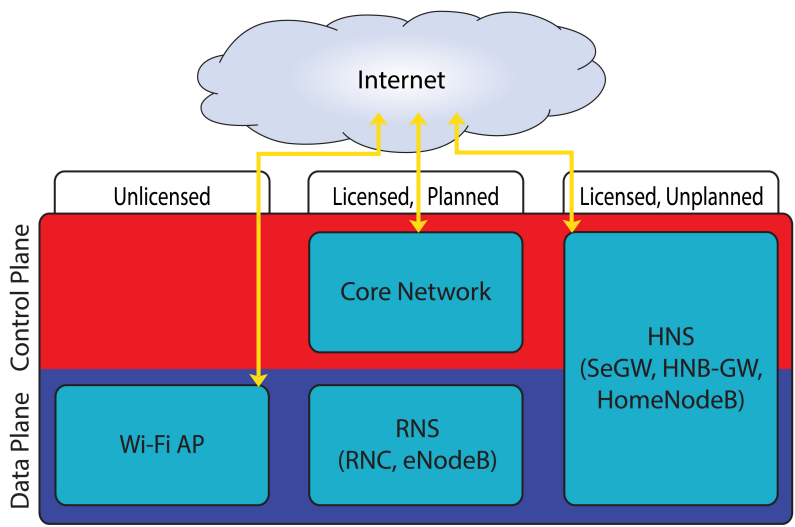

Figure 3: Co-existence of heterogeneous networks may be managed in part by clients

*Case 4: Client-controlled Cloud storage (in beta trial).* By decoupling massive cheap storage (in the Cloud) from client side control of privacy (in the Fog), we can achieve the best of both worlds. For example, by spreading the bytes, in a client shim layer, of a given file across multiple Cloud storage providers, it can be assured that privacy of the data is maintained even if encryption key is leaked by any given Cloud provider [5].

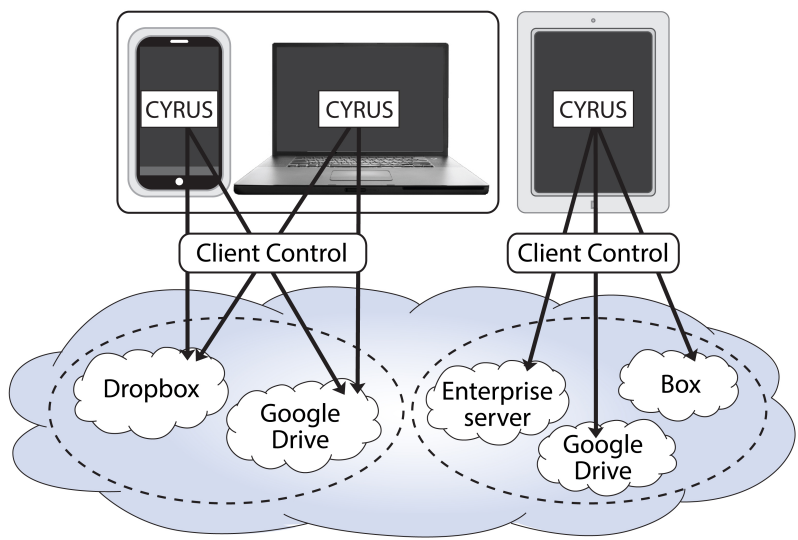

Figure 4: Shred and Spread (CYRUS project) stores in Cloud but controls in Fog

*Case 5: Real-time stream mining (in beta trial).* Consider virtual reality tasks associated with Google Glass. Some of the information retrieval and computation tasks may be carried out on the Glass (a "wearable thing"), some on the associated phone (a client device), some on the home storage (an edge device), and the rest in the Cloud. An architecture of successive refinement may leverage all of these devices at the same time, with an intelligent division of labor across them [6].

*Case 6: Borrowing bandwidth from neighbors in D4D (in beta trial).* When multiple devices belonging to the same person, to relatives or to employees of the same company are next to each other, one can ask the others to share their LTE/WiFi bandwidth by downloading other parts of the same file and transmitting, via WiFi Direct, client to client [7].

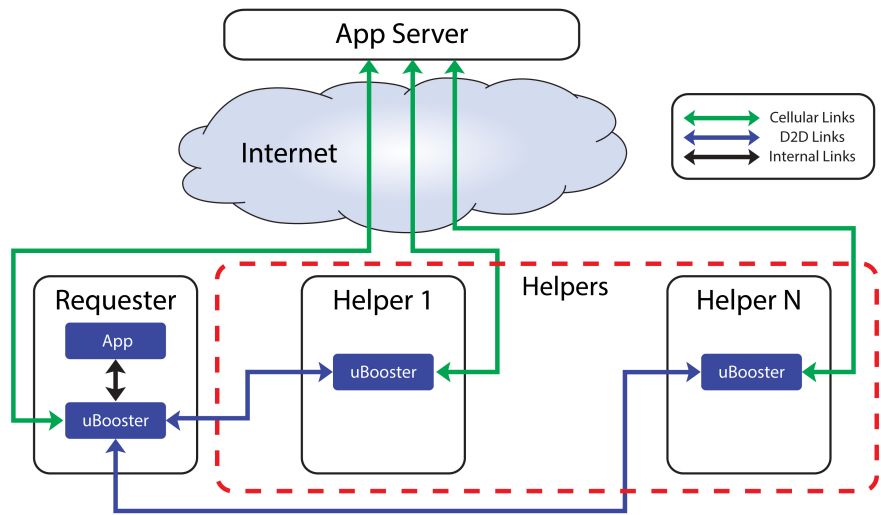

Figure 5: Idle resources in client devices can be pooled in D4D for more efficient use

*Case 7: Bandwidth management at home gateway (in beta trial)*. By adapting the home set-top box/gateway, the limited broadband capacity is allocated among competing users and application sessions, according to each session's priority and individual preferences. A prototype on a commodity router demonstrates a scalable, economical and accurate control of capacity allocation on the edge [8].

*Case 8: Distributed beam-forming (in lab demonstration)*. Fog can also happen in the physical layer, for example, by exploiting multi-user MIMO to improve throughput and reliability when a client can communicate with multiple WiFi access points. For uplink, we can use multi-user beam-forming so that the client can send multiple data streams to multiple APs simultaneously. For downlink, we can use interference nulling so that the client can decode parallel packets from multiple APs. These can be done entirely on the client side [9].

Some of the above case studies are core topics in what many people imagine would partially define "5G:" HetNets/small cell/densification, over the top service provisioning, cognitive radio and crowd-sensing. Other case studies point toward architectural thinking for IoT services, questions about ownership, control and visibility of personal area networks, such as "should Apple Watch and the like have their own data plan?" that will help define the balance of power between "AT&T"s and "Apple"s of the world. If the network in or around the end users have a logical topology that looks like a star, with a fixed gateway (e.g., iPhone), the visibility, control, and value-added by network operators will be drastically different than in the alternative scenario where the gateways are dynamically chosen or the Things can sometimes have direct communication paths without a gateway.

For more references for these examples and more examples, please see an initial list of over 100 recent publications on eight different topics under Fog Networking at http://fogresearch.org

**D. Open Questions and Research Challenges**

As is typical of any emergent area of R&D, many of the themes in Fog Networking are not completely new, and instead are evolved version of accumulated transformations in the past decade or two:

- Compared to peer-to-peer (P2P) networks in the mid-2000s, Fog is not just about content sharing (or data plane as a whole), but also network measurement, control & configuration, and service definition.
- Compared to mobile ad hoc network (MANET) research a decade ago, we have much more powerful and diverse off-the-shelf edge devices and applications now, together with the structure/hierarchy that comes with cellular/broadband networks.
- Compared to generic edge-networking in the past, Fog networking provides a new layer of meaning to the end-to-end principle: not only do edge devices optimize among themselves, but they collectively measure and control the rest of the network.

Along with two other network architecture themes: ICN and SDN, each with a longer history, Fog is revisiting the foundation of how to think about and engineer networks, i.e., how to *optimize network functions:* who does what and how to glue them back together:
- Information-Centric Networks: *Redefine* functions (to operate on digital objects rather than just bytes)
- Software-Defined Networks: *Virtualize* functions (through centralized control plane)
- Fog Networks: *Relocate* functions (to the network edge)

While Fog networks do not have to have any virtualization or to be information-centric, one could also imagine an information-centric, software-defined Fog network (since these three branches are not orthogonal).

As in an emergent area in its infant age, there is no shortage of challenging questions in Fog networking, some of which continue from earlier study of P2P, MANET and Cloud, while others are driven by a confluence of recent developments in network engineering, device systems and user experience:

- *Cloud-Fog interface*: The fundamental question of architecture is "who does what, at what timescale, and how to put them back together?" Cloud will remain useful as Fog arises. The question is what tasks go to Fog (e.g., those that require real-time processing, end user objectives or low-cost leverage of idle resources) and what go to Cloud (e.g., massive storage, heavy-duty computation, or wide-area connectivity), and what will be the Fog-Cloud and Fog-Fog interfaces: the specification of information passage, from its frequency to granularity, across these interfaces.

- *Interactions with client/thing hardware and OS*: Once the actions are taken on the clients or things, the interface with their operating systems and hardware become essential. More than just using D4D for pooling idle edge resources, there is also the possibility of specialized protocol stack just for networking within an edge.

- *Trustworthiness and security*: While Fog may help enhance security in some cases, it may present new security challenges in other cases. Given that it is often easier to hack into client software, perhaps security at hardware level on client devices. At the same time, because of the proximity to end users and locality on the edge, nodes in Fog networks can often act as the first node of access control and encryption, provide contextual integrity and isolation, and enable the control of aggregating privacy-sensitive data before it leaves the edge.

- *Incentivization of client participation*: Sometimes it is not too many un-trustworthy clients that create concern but too few clients willing to participate. Market systems and incentive mechanisms will become useful.

- *Convergence and consistency* arising out of local interactions: Typical concerns of distributed control, divergence/oscillation and inconsistency of global states, become more acute in a massive, under-organized, possibly mobile crowd with diverse capabilities and virtualized pool of resources shared unpredictably. Use cases in edge analytics and stream mining provide additional challenges on this recurrent challenge in distributed systems.

- In general, the tradeoff between *distributed and centralized* architectures, between what stays on local and what goes on global, and between careful planning and resilience through redundancy. On this topic, we need to be sensitive to the opportunity where many different, or dynamic, logical topologies may arise from the same underlying physical configuration of a Fog network.

To address the above challenges, we need both
- Fundamental research, across networking, device hardware/OS, pricing, HCI and data science, and
- Industry-academia interactions, as exemplified in the Open Fog Consortium, a global, non-profit consortium launched in November 2015 with founding members from ARM, Cisco, Dell, Intel, Microsoft and Princeton University.

Indeed, Fog Networking is starting to shape the future of the balance of power and distribution of driving innovation across the entire industry food chain, including the following:

- End user experience provider (e.g., GE, Toyota…)
- Network operators (e.g., AT&T, Verizon, Comcast…)
- Network equipment vendors (e.g., Cisco, Nokia, Ericsson, Huawei…)
- Cloud service providers (e.g., VMWare, Amazon…)
- System integrators (e.g., IBM, HP…)
- Edge device manufacturers (e.g., Linksys…)
- Client/IoT device manufacturers (e.g., Dell, Microsoft, Apple, Google…)
- Chip suppliers (e.g., Intel, ARM, Qualcomm, Broadcom…)

2015 is an interesting year to start systematically exploring what Fog might look like and the differences it will make in the world of networking and computing in the next 15 years.

**Acknowledgements:**

The author is grateful for the inspiring conversations with many colleagues in industry and academia, especially Flavio Bonomi, Russell Hsing, Bharath Balasubramanian, Aakanksha Chowdhery, Yan Schvartzshnaider, Sangtae Ha, Junshan Zhang, Raj Savoor, John Smee, Chonggang Wang and representatives of ARM, Cisco, Dell, Intel and Microsoft in Open Fog Consortium .